\documentclass{webofc}
\usepackage[varg]{txfonts}   
\begin{document}
\title{Dynamics of bubble walls at the electroweak \\phase transition}
%
%

\author{\firstname{Stefania} \lastname{De Curtis}\inst{1}\fnsep\thanks{\email{stefania.decurtis@fi.infn.it}} \and
        \firstname{Luigi} \lastname{Delle Rose}\inst{2}\fnsep\thanks{\email{luigi.dellerose@unical.it}} \and
        \firstname{Andrea} \lastname{Guiggiani}\inst{1}\fnsep\thanks{\email{andrea.guiggiani@unifi.it}} \and
        \firstname{\'Angel} \lastname{Gil Muyor}\inst{3}\fnsep\thanks{\email{agil@ifae.es}} \and
        \firstname{Giuliano} \lastname{Panico}\inst{1}\fnsep\thanks{\email{giuliano.panico@unifi.it}} 
}

\institute{INFN Sezione di Firenze and Dipartimento di Fisica e Astronomia, Universit\`a di Firenze, Via G. Sansone 1, I-50019 Sesto Fiorentino, Italy
\and
           Dipartimento di Fisica, Universit`a della Calabria, I-87036 Arcavacata di Rende, Cosenza, Italy
\and
           IFAE and BIST, Universitat Aut\`onoma de Barcelona, 08193~Bellaterra,~Barcelona,~Spain
          }

\abstract{%
First order phase transitions in the early universe naturally lead to the production of a stochastic background of gravitational waves and to the generation of a matter-antimatter asymmetry.
The dynamics of the phase transition is affected by the density perturbations in the hot plasma. 
We address this topic by providing, for the first time, a full numerical solution to the linearized Boltzmann equation for the top quark species coupled to the Higgs field during a first order phase transition at the electroweak scale. Differently from the traditional approaches, our results do not depend on any ansatz and can fully describe the non-equilibrium distribution functions of the particle species in the plasma.
}
\maketitle

\section{Introduction}\label{sec:intro}

The recent observation of gravitational waves has renewed a vivid interest in the study of the dynamics of first order phase transitions. 
Indeed, future gravitational wave experiments, such as the European proposal LISA~\cite{Caprini:2015zlo,Caprini:2019egz}, the Japanese interferometer DECIGO~\cite{Kawamura:2006up,Kawamura:2011zz} and the Chinese projects Taiji~\cite{Hu:2017mde,Ruan:2018tsw} and TianQin~\cite{TianQin:2015yph} will probe a range of the expected peak frequencies of stochastic background of gravitational waves sourced by PhTs at the electroweak scale. These interferometers will provide us with a new handle that, supporting already existing collider experiments, will help us in the quest for the physics beyond the Standard Model (BSM), and in particular for all the models affecting the dynamics of the electroweak (EW) symmetry breaking.

It is worth mentioning that the stochastic gravitational wave background is not the only cosmological relic left after the completion of a PhT. Indeed, the same mechanism can also produce dark matter remnants, an asymmetry in the matter-antimatter abundances, primordial black holes, magnetic fields and other topological defects.
Obviously, a reliable characterization of these processes requires an accurate modelling of the PhT dynamics,
which is controlled, among other parameters, by the propagation speed of the PhT front, namely, the bubble wall. 

In the steady state regime, the velocity of the bubble wall is a result of a balance of the external friction exerted by the particles in the plasma impinging on the wall and of the internal pressure, caused by the potential difference between the two phases. 
In fact, the plasma is driven out of equilibrium by the presence of the moving bubble and induces a backreaction that slows down its propagation.
This is an extremely relevant topic, as shown by the huge amount of literature, for an incomplete list see for instance~\cite{Moore:1995ua,Moore:1995si,John:2000zq,Moore:2000wx,Konstandin:2014zta,Kozaczuk:2015owa,Bodeker:2017cim,Cline:2020jre,Laurent:2020gpg,BarrosoMancha:2020fay,Hoche:2020ysm,Azatov:2020ufh,Balaji:2020yrx,Cai:2020djd,Wang:2020zlf,Friedlander:2020tnq,Cline:2021iff,Cline:2021dkf,Bigazzi:2021ucw,Ai:2021kak,Lewicki:2021pgr,Gouttenoire:2021kjv,Dorsch:2021ubz,Dorsch:2021nje,Megevand:2009gh,Espinosa:2010hh,Leitao:2010yw,Megevand:2013hwa,Huber:2013kj,Megevand:2013yua,Leitao:2014pda,Megevand:2014yua,Megevand:2014dua}, nevertheless, the speed of the wall is one of the parameters characterizing the PhT dynamics on which we have less theoretical control.

The first computation of the bubble speed in the SM has been carried out by Moore and Prokopec in their seminal work~\cite{Moore:1995ua,Moore:1995si}. 
The friction on the bubble wall exerted by the plasma was calculated by evaluating all the relevant interactions between the plasma particles and the bubble and
required the solution of the Boltzmann equation to determine the out of equilibrium distribution functions of the different species in the plasma through.
Besides this microphysical calculation, other phenomenological approaches have also been explored. These relied, instead, on a modelling of the friction and on its parameterization in terms of the viscosity parameter~\cite{Megevand:2009gh,Espinosa:2010hh,Leitao:2010yw,Megevand:2013hwa,Huber:2013kj,Megevand:2013yua,Leitao:2014pda,Megevand:2014yua,Megevand:2014dua}.

At the core of the formalism developed in~\cite{Moore:1995ua,Moore:1995si}, that in the following we will denote as the ``old formalism'', there is the exploitation of an ansatz for the distribution functions.
This is necessarily needed in order to parametrize the momentum dependence of the out of equilibrium distributions and to compute the collision integral appearing in the Boltzmann equation. 
In particular, in ref.~\cite{Moore:1995si} the fluid approximation was chosen as an ansatz: it assumes that the out of equilibrium distribution function of each species in the plasma is entirely described by only three perturbations: the chemical potential, the temperature fluctuation and the velocity perturbation. These corrections are extracted by taking suitable moments of the Boltzmann equation with specific weights. With such a method, the integro-differential Boltzmann equation is recast into a much more manageable system of ordinary differential equations.

Two the peculiar features of the fluid approximation explored in~\cite{Moore:1995si} are: a) the Liouville operator of the Boltzmann equation develops a zero eigenvalue at the speed of sound $c_s$ and, b) for velocities larger than $c_s$ all the perturbations trail the source term. This has a large impact on models of non-local EW baryogenesis because any non-equilibrium dynamics in front of the bubble wall would be suppressed for bubble walls faster than $c_s$~\cite{Konstandin:2014zta}, resulting to a very inefficient production of the matter-antimatter asymmetry. Recently in refs.~\cite{Cline:2020jre,Laurent:2020gpg} it has been suggested that the singular behavior at the speed of sound is only an artifact of the truncation in momenta introduced by the fluid approximation and of the corresponding weights used to compute the out-of-equilibrium perturbations. Indeed, changing the set of weights can modify the location of the singularity, which clearly suggests that the speed of sound cannot be a critical value that characterizes particle diffusion in the fluid equations. In ref.~\cite{Laurent:2020gpg} the authors overcame this problem by developing a ``new formalism'' that relies on a different parameterization of the non-equilibrium distributions, different weights as well as on a factorization ansatz~\cite{Cline:2000nw}. The new formalism erases the discontinuity at the speed of sound and still provides results quantitatively similar to those of the fluid approximation for small enough velocities. 

The same topic has been recently explored also in ref.~\cite{Dorsch:2021ubz}, specifically for the calculation of the matter-antimatter asymmetry, and in ref.~\cite{Dorsch:2021nje} for the calculation of the friction on the bubble wall. In these works the fluid approximation of ref.~\cite{Moore:1995si} has been generalized by including higher orders in the small momenta expansion
and the absence of the singularity for the out-of-equilibrium distribution functions of the heavy species has been confirmed (notice that the fluid approximation can be seen as a first order expansion in the momenta).  Besides that, the speed of sound in the plasma still played a peculiar role in the old formalism (even if higher orders in the momenta expansion are included, assuming that the thickness of the wall is not too large and the interaction strength among the plasma particles is not too strong) as it provides peaks in the integrated friction, one for each vanishing eigenvalue of the Liouville operator.
For the massless background species, instead, as already pointed out in ref.~\cite{Moore:1995si} and confirmed in ref.~\cite{Dorsch:2021nje,Laine:1993ey,Ignatius:1993qn,Kurki-Suonio:1995rrv,Espinosa:2010hh}, a discontinuity at $v_w \simeq c_s$ of the out-of-equilibrium perturbations in the temperature and the velocity profiles of the fluid could be present. This discontinuity manifests itself in a singularity at $c_s$ if the Boltzmann equations are linearized. \\
Besides the singularity, large corrections in the friction, with respect to the old formalism, have also been pointed out in~\cite{Dorsch:2021ubz,Dorsch:2021nje} when higher orders in momenta are taken into account. These results confirms that the fluid approximation cannot provide reliable predictions, neither quantitatively nor qualitatively.

The methodologies discussed above are affected by two main drawbacks: first of all, they must rely on an ansatz describing the dependence in the momenta of the non-equilibrium distribution functions. This is unavoidable to allow for the computation the collision integrals; secondly, the choice of the momentum basis used in the ansatz and the choice of the weights used to construct the system of ordinary differential equations are not unique: different ansatzes have qualitative and quantitative impacts on the resulting distribution functions. 
This clearly indicates that a full solution to the Boltzmann equation that does not impose any specific momentum dependence
is necessary to provide reliable quantitative predictions for both the non-equilibrium distribution functions and the friction exerted on the bubble wall. As a byproduct this will also allow to clarify the issue of the presence of a singularity. 
A first step in this direction has been carried out in~\cite{DeCurtis:2022hlx}, where we have presented, for the first time, a fully quantitative numerical solution to the Boltzmann equation. For the purpose of presenting the methodology and to quantitatively asses the differences among the aforementioned formalisms, we have considered the EWPhT and we have focused on the study of top quark species, the one that provides the largest contribution to the friction.
By feeding the non-equilibrium distribution functions into the equation of motion of the Higgs field, one will be able to carry out a precise determination of the bubble profile and of the the wall velocity. As we have already pointed at the beginning of the introduction, this is necessary in order to clearly identify, at the quantitative level, the potential of a given BSM scenario to yield interesting predictions for the cosmological relics listed above.

\section{The Boltzmann equation}\label{sec:Boltzmann}

The aim is to find the solution to the Boltzmann equation for the distribution function of
the plasma species in the presence of a bubble wall expanding in the true vacuum. 
For large enough bubbles we can adopt the planar limit and we can assume a steady state regime. In this case the solutions to the Boltzmann equation 
are stationary in the wall frame. Orienting the $z$ axis along the velocity of the
bubble wall, the equation for the distribution function $f$ of a particle species is 
\begin{equation}
{\cal L}[f] \equiv \left(\frac{p_z}{E} \partial_z - \frac{(m^2(z))'}{2E} \partial_{p_z}\right) f = - {\cal C}[f]\,,
\end{equation}
where $m(z)$ is the position-dependent mass of the particle. The term ${\cal C}$ is the collision integral, describing local microscopic
interactions among the plasma particles, while ${\cal L}$ is the Liouville operator.

The collision term ensures that far from the wall each particle species approaches
local thermal equilibrium described by the standard Fermi or Bose--Einstein distributions for a fluid moving with velocity $v$ along the
$z$ axis, namely
\begin{equation}
f_v = \frac{1}{e^{\beta \gamma(E - v p_z)} \pm 1}\,,
\end{equation}
with $\beta = 1/T$ and $\gamma = 1/\sqrt{1- v^2}$.

Deviations with respect to the local equilibrium distribution are present close to the wall and vanish at $z \to \pm \infty$. 
For small enough perturbations, the distribution function can be written as
$f = f_v + \delta f$ and the Boltzmann equation can be linearized in $\delta f$:
\begin{equation}\label{eq:Boltz_lin}
\left(\frac{p_z}{E} \partial_z - \frac{(m(z)^2)'}{2E} \partial_{p_z}\right) \delta f + {\overline{\cal C}}[\delta f] = \frac{(m(z)^2)'}{2E} \partial_{p_z} f_v = \beta \gamma v \frac{(m(z)^2)'}{2E} f_v'\,,
\end{equation}
where we defined
\begin{equation}
f_v' \equiv - \frac{e^{\beta \gamma(E- v p_z)}}{(e^{\beta \gamma(E- v p_z)} \pm 1)^2}\,,
\end{equation}
and ${\overline{\cal C}}[\delta f]$ denotes the collision integral linearized in $\delta f$.
Sizable values for the source term in the equation above are present only close to the wall, where the non-trivial Higgs profile
generates a non-negligible $z$ dependence. Away from the wall, where the Higgs profile is almost constant, the source term is suppressed, in agreement with the expectation that deviations from thermal equilibrium are only present close to the wall and should decrease to zero away from it.

\subsection{The Liouville operator and the collision integral}\label{sec:Liouville}

Along the paths on which both the transverse momentum $p_\bot$ and the quantity $p_z^2 + m^2(z)$ are constant, the Liouville differential operator reduces to a total derivative with respect to $z$:
\begin{equation}
{\cal L} = \left(\frac{p_z}{E} \partial_z - \frac{(m^2(z))'}{2E} \partial_{p_z}\right) \quad \to \quad \frac{p_z}{E} \frac{d}{dz}\,.
\end{equation}

\begin{figure}
\centering
\includegraphics[width=.52\textwidth]{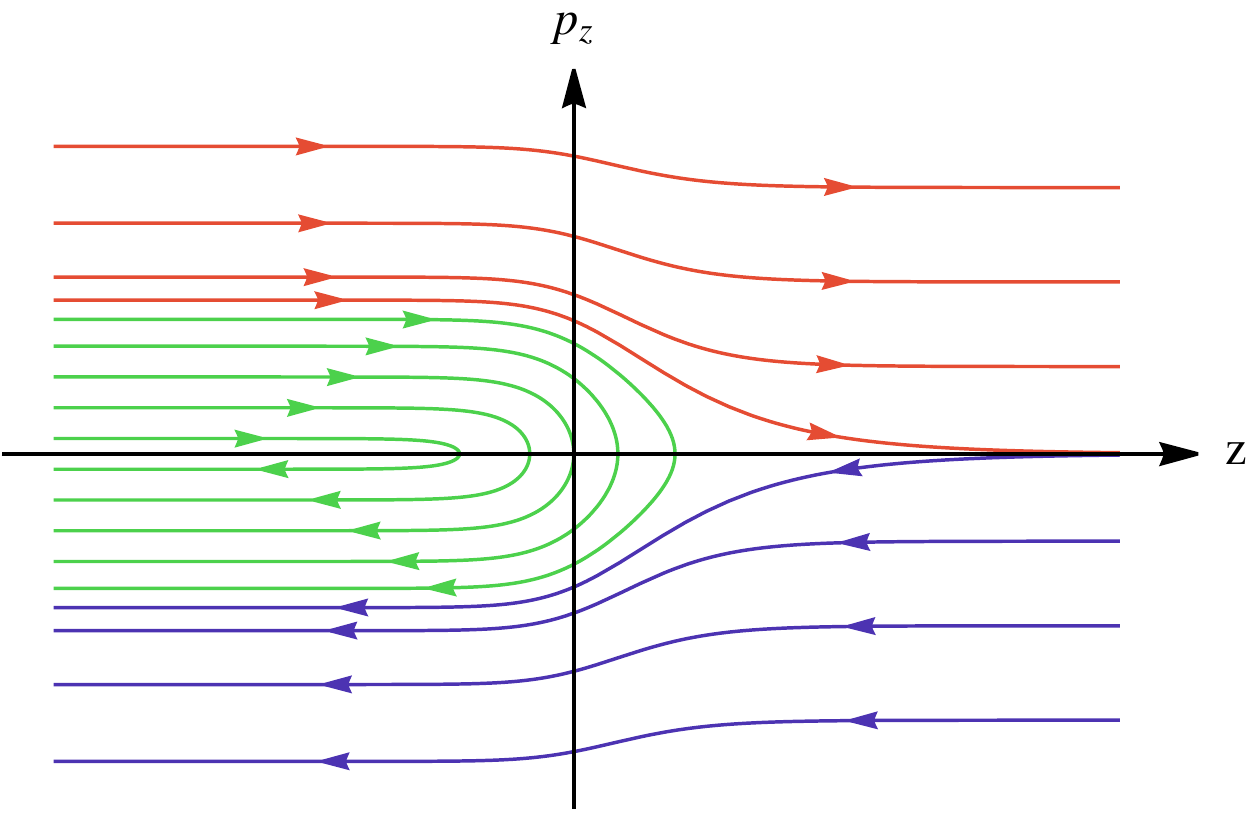}
\hfill
\raisebox{1em}{\includegraphics[width=.43\textwidth]{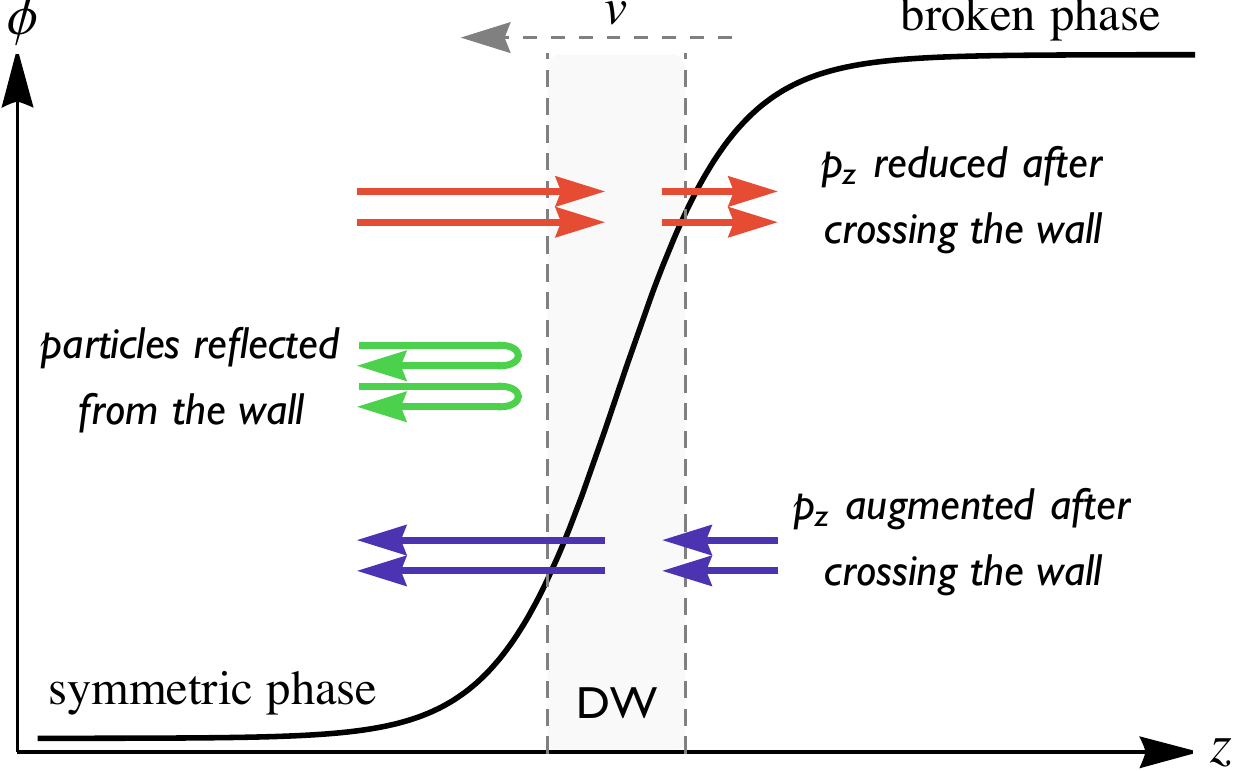}}
\caption{Left panel: Paths with fixed energy and transverse momentum in the $z - p_z$ phase space for the choice $m(z) \propto 1 + \tanh(z/L)$. The red, green and purple colors denote sets of contours with different behavior. The arrows show the flow of
a particle within the phase space. Right panel: Schematic representation of the behavior of the particles across the bubble wall.}\label{fig:flow_paths}
\end{figure}

The paths represent the trajectories of the particles in the $(p_\bot, p_z, z)$ phase space in the collisionless limit in which the energy and the momentum parallel to the wall are conserved.
As such the trajectories of the particles are given by the condition $p_z^2 + m^2(z) = const$ where $m(z) \to 0$ for $z \to -\infty$,
and $m(z) \to m_0 > 0$ for $z \to +\infty$.
The three classes of curves are shown schematically in fig.~\ref{fig:flow_paths} for the choice $m(z) \propto 1 + \tanh(z/L)$, with $L$ denoting the wall thickness.
The first type of paths describes particles that travel in the positive $z$ direction, and eventually enter into the bubble. 
The second type of paths describes particles that initially travel in the positive $z$ direction, hit the wall and are reflected.
The third type of paths describes particles that travel in the negative $z$ direction, and eventually exit from the bubble.  \\

The collision integral for the $2 \to 2$ processes is given by
\begin{equation}
{\cal C}[f] = \sum_i \frac{1}{4N_p E_p} \int\! \frac{d^3{\bf k}\,d^3{\bf p'}\,d^3{\bf k'}}{(2\pi)^5 2 E_k 2E_{p'} 2E_{k'}} |{\cal M}_i|^2
\delta^4(p+k-p'-k') {\cal P}[f]\,,
\end{equation}
with the population factor
\begin{equation}
{\cal P}[f] = f(p) f(k)(1 \pm f(p'))(1 \pm f(k')) - f(p') f(k')(1 \pm f(p))(1 \pm f(k))\,,
\end{equation}
where the sum is over all the scattering processes described by a squared scattering amplitude $|{\cal M}_i|^2$.
Moreover, $N_p$ is the number of degrees of freedom of the incoming particle with momentum $p$, $k$ is the momentum of the second incoming particle, while $p'$ and $k'$ are the momenta of the outgoing particles. The $\pm$ signs are $+$ for bosons and $-$ for fermions.

\section{Numerical analysis}\label{sec:numerical}

To solve the Boltzmann equation we have developed a numerical iterative method described in~\cite{DeCurtis:2022hlx}.
For the sake of simplicity, we considered a single species in the plasma, the top quark, which is expected to provide the most relevant effects to the bubble dynamics, being the state with largest coupling to the Higgs. The main contribution to the collision integral comes from the annihilation process $t \bar t \to g g$, whereas the scattering of tops on gluons and light quarks gives smaller contributions. \\
The bubble wall is modelled assuming that the Higgs profile can be approximated by
\begin{equation}
\phi(z) = \frac{\phi_0}{2}[1 + \tanh(z/L)]\,,
\end{equation}
where $L = 5/T$ is the thickness of the bubble wall and  $\phi_0 = 150\;{\rm GeV}$ is the Higgs VEV in the broken phase. We fixed the
phase transition temperature to $T = 100\;{\rm GeV}$. \\
An important quantity that can be obtained from the numerical solution is the friction acting on the bubble wall~\cite{Moore:1995si}
\begin{equation}
F(z) = \frac{d m^2}{d z} N \int \frac{d^3 {\bf p}}{(2 \pi)^3 2E} \delta f(p)\,,
\end{equation}
where $N$ denotes the number of degrees of freedom.

\begin{figure}
	\centering
	{\includegraphics[width=.47\textwidth]{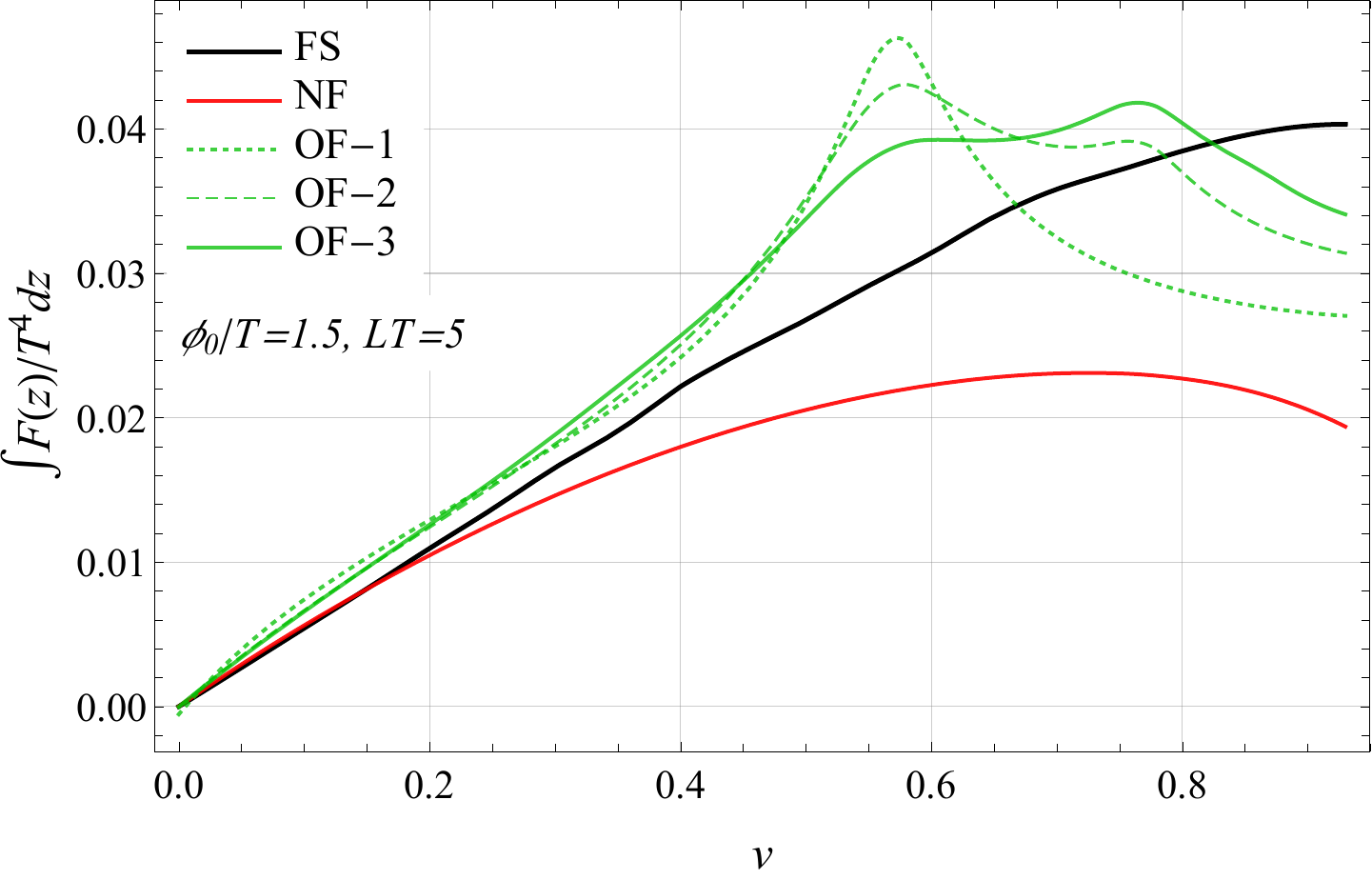}}
	\caption{Friction acting on the bubble wall as a function of the velocity. The black solid line corresponds to the solution of the full Boltzmann equation (FS, our result), the dotted, dashed and solid green lines are obtained with the old formalism (OF) at order $1$, $2$ and $3$ respectively~\cite{Moore:1995si,Dorsch:2021ubz}, while the red line corresponds to the new formalism (NF)~\cite{Laurent:2020gpg}.}\label{fig:friction}
\end{figure}

In fig.~\ref{fig:friction} we show the friction integrated over $z$ as a function of the wall velocity (solid black line).
We also compare our result (full solution, FS) with the ones obtained with the weighted methods.
The green lines correspond to the total friction computed in the old formalism (OF) of ref.~\cite{Moore:1995si},
taking also into account higher-order corrections ($1$, $2$ and $3$) to the fluid approximation~\cite{Dorsch:2021ubz}. The red line represents, instead, the result obtained in
new formalism (NF) of ref.~\cite{Laurent:2020gpg}. \\
For small and intermediate velocities, $v \lesssim 0.3$, the old formalism is in a fair numerical agreement with the full solution.  
At higher velocities, instead, the old formalism develops peaks corresponding to the zero eigenvalues of the Liouville operator and
highlighting an instability in the methodology. Our results for the full solution to the Boltzmann equation confirm that the peaks are an artifact and that no strong effect is present  close to the speed of sound if the top contributions are taken into account. \\
Concerning the new formalism, it correctly predicts a smooth behavior for the total friction for all the velocities but the quantitative agreement with the full solution is good only for very low ones.

\section{Conclusions and Outlook}\label{sec:conclusions}

In~\cite{DeCurtis:2022hlx} we presented for the first time the fully quantitative solution of the Boltzmann equation that describes the distribution functions of plasma particles in the presence of an expanding bubble. 
Differently to the existing approaches, we did not exploited any ansatz nor we imposed any momentum dependence on the distribution functions away from equilibrium.
We have then critically compared our results with the ones obtained using the formalisms developed so far in the literature, namely the fluid approximation~\cite{Moore:1995si}, its extended version~\cite{Dorsch:2021ubz,Dorsch:2021nje} and the ``new formalism''~\cite{Laurent:2020gpg}. \\
This clearly represents a necessary step towards a reliable understanding of the bubble wall dynamics. 
Using the friction obtained with the numerical method developed in our work~\cite{DeCurtis:2022hlx}, one can solve the equation of motion of the Higgs profile and extract the velocity of the bubble wall as well as the features of its shape, such as the wall thickness. These parameters crucially impact on the prospects of any BSM scenario to predict cosmological signals, such as a stochastic gravitational wave background and the amount of matter-antimatter asymmetry.

\end{document}